\documentclass[11pt]{article}
\newcommand{\Comment}[1]{{}}
\usepackage{amsfonts,amsthm,amsmath,amssymb,slashed}
\usepackage[textwidth = 430 pt, textheight = 630 pt]{geometry}

\Comment{\usepackage{color}
\definecolor{MyDarkBlue}{rgb}{0.15,0.15,0.45}
\usepackage[linktocpage=true]{hyperref}
\hypersetup{
colorlinks=true,
citecolor=MyDarkBlue,
linkcolor=MyDarkBlue,
urlcolor=MyDarkBlue,
pdfauthor={Horatiu Nastase and Amanda Weltman},
pdftitle={Chameleons on the racetrack},
pdfsubject={hep-th}
}

\usepackage[numbers,sort&compress]{natbib}
\usepackage{hypernat}}
\usepackage{graphicx}
\usepackage{cite}

\newcommand\ignore[1]{}
\def\one{{\,\hbox{1\kern-.8mm l}}}

\def\d{\partial}

\def\lsim{\mathrel{\mathstrut\smash{\ooalign{\raise2.5pt\hbox{$<$}\cr\lower2.5pt\hbox{$\sim$}}}}}
\def\gsim{\mathrel{\mathstrut\smash{\ooalign{\raise2.5pt\hbox{$>$}\cr\lower2.5pt\hbox{$\sim$}}}}}

\newcommand{\be}{\begin{equation}}
\newcommand{\bea}{\begin{eqnarray}}

\newcommand{\ee}{\end{equation}}
\newcommand{\eea}{\end{eqnarray}}

\begin{document}

\renewcommand{\thefootnote}{\fnsymbol{footnote}}

\makeatletter
\@addtoreset{equation}{section}
\makeatother
\renewcommand{\theequation}{\thesection.\arabic{equation}}

\rightline{}
\rightline{}
   \vspace{1.8truecm}


\vspace{10pt}


\begin{center}
{\LARGE \bf{\sc Chameleons on the Racetrack}}
\end{center} 
 \vspace{1truecm}
\thispagestyle{empty} \centerline{
{\large \bf {\sc Horatiu Nastase${}^{a,}$}}\footnote{E-mail address: \Comment{\href{mailto:nastase@ift.unesp.br}}{\tt nastase@ift.unesp.br}}
{\bf{\sc and}} 
{\large \bf {\sc Amanda Weltman${}^{b,}$}}\footnote{E-mail address: \Comment{\href{mailto:amanda.weltman@uct.ac.za}}{\tt amanda.weltman@uct.ac.za}}
                                                           }

\vspace{.8cm}

\centerline{{\it ${}^a$ 
Instituto de F\'{i}sica Te\'{o}rica, UNESP-Universidade Estadual Paulista}} \centerline{{\it 
R. Dr. Bento T. Ferraz 271, Bl. II, Sao Paulo 01140-070, SP, Brazil}}

\vspace{.7cm}

\centerline{{\it ${}^b$ Astrophysics, Cosmology \& Gravity Center,}}
\centerline{{\it Department of Mathematics and Applied Mathematics, 
University of Cape Town}} 
\centerline{{\it Private Bag, Rondebosch 7700,  South Africa}}

\vspace{2truecm}

\thispagestyle{empty}

\centerline{\sc Abstract}

\vspace{.4truecm}

\begin{center}
\begin{minipage}[c]{380pt}
{\noindent We modify the ansatz for embedding chameleon scalars in string theory proposed in \cite{Hinterbichler:2010wu} by considering a racetrack
superpotential with two KKLT-type exponentials $e^{ia\varrho}$ instead of one. This satisfies all experimental constraints, while also allowing for 
the chameleon to be light enough on cosmological scales to be phenomenologically interesting.}
\end{minipage}
\end{center}

\vspace{.5cm}

\setcounter{page}{0}
\setcounter{tocdepth}{2}

\newpage

\renewcommand{\thefootnote}{\arabic{footnote}}
\setcounter{footnote}{0}

\linespread{1.1}
\parskip 4pt


\section{Introduction}
\ \ \ \ \

Both in phenomenological and effective field theories, as well as in fundamental theories like string theory, we have generically one or many scalar fields, 
usually of small mass, that in principle could play a role in cosmology. Generically this is a problem however, since we have not yet observed such scalars.\footnote{We have 
just observed a scalar, likely the Higgs, at the LHC \cite{:2012gk,:2012gu}, 
but that is very massive, and it is very hard to make a model where the Higgs is a scalar relevant 
for cosmology, like the inflaton. It is also not clear as of yet if this scalar is fundamental, or is some composite object which was not present at early 
times.} Indeed, there is usually an argument against light scalar fields as they produce a fifth force, violate the equivalence principle and thus disturb the predictions of general relativity at planetary or solar system scales, while also affecting laboratory experiments on Earth. 

The idea of chameleon scalars \cite{Khoury:2003aq,Khoury:2003rn} was introduced as a way to avoid these constraints, while still having a light scalar 
on cosmological scales \cite{Brax:2004qh,Brax:2004px}. A chameleon scalar has an effective potential, and in particular a mass, depending on the local matter density. As a result,
on solar system and planetary scales, the constraints are satisfied because the chameleon is screened, only a thin shell around a large spherical body effectively 
interacts via the scalar, whereas on Earth the lab constraints are evaded because the mass of the scalar on Earth is large, due to the large ambient densities 
(Earth and atmosphere). However, it has proven challenging to embed the chameleon mechanism inside a fundamental theory. In particular, in string theory
we have a large number of scalar moduli, generically light, coming among others from the size and shape parameters of the compact space. Usually we have to find 
a method to stabilize these moduli, i.e. to give them large masses around a minimum. This is a notoriously difficult problem
\cite{Damour:1994zq,Damour:1994ya}, which would be alleviated if we
could have one or more of these moduli be a chameleon, hence the increased interest in finding an embedding of the chameleon idea inside string theory.

In \cite{Hinterbichler:2010wu}, a possible way to do this was proposed, where the chameleon scalar is the volume modulus $\varrho$ 
for the compact space. A general phenomenological way to obtain a chameleon theory based on a supergravity compactification was proposed, with a potential for 
the volume modulus with a quadratic approximation around a stabilized minimum, together with a steep exponential on the large volume side of the potential. 
An example was given, based on the KKLT construction \cite{Kachru:2003aw}, but where the KKLT superpotential $W=W_0+Ae^{ia\varrho}$ has $a<0$ instead of 
$a>0$.\footnote{This is however possible, as stressed in \cite{Hinterbichler:2010wu}, even in the context of KKLT \cite{Abe:2005rx}, as well as
in more general string contexts \cite{Quevedo:1996sv,Burgess:1997pj}.} In \cite{Hinterbichler:2010wu} it was also assumed that $\varrho$ was in units of 
four dimensional $M_{\rm Pl}$, as opposed to fundamental string units (related to 10 dimensional Planck scale), which implies that $a$ must be much larger
than its natural value. This problem was eliminated in \cite{HNsoon}, where $\varrho$ was assumed to be in fundamental string units, being forced by experimental
constraints to have a scenario with two large and varying extra dimensions, and the other four fixed. The KKLT potential was not a perfect example of the 
general phenomenological case, the only important difference being that it led to a chameleon mass on cosmological scales constrained by $m_{\rm cosmo}\gsim
10^{15}H_0$ (as opposed to $m_{\rm cosmo}\gsim 10^3H_0$ for the general phenomenological potential), which makes it less interesting for cosmology. 

In this note, by considering a racetrack potential (for uses and abuses see \cite{Denef:2004dm, BlancoPillado:2004ns, Greene:2005rn}), we show that a simple modification of the model solves the problem. Specifically, by adding two KKLT-type exponentials instead of one we obtain a 
model that still satisfies all experimental constraints, while being interesting for cosmology. In section 2 we present the model, first reviewing the 
set-up in \cite{Hinterbichler:2010wu,HNsoon} and then modifying it for our purposes. In section 3 we check the experimental constraints, and verify that there exist parameters that satisfy them. This is a proof in principle that this can be done, however we do not claim to solve any of the fine tuning problems associated either with racetrack potentials or the cosmological constant problem. 

\section{The model}

When dimensionally reducing a 10-dimensional gravitational theory down to four dimensions, in general we make a reduction ansatz
\bea
\nonumber
{\rm d}s^2_D&=&R^2{\rm d}s_4^2+g_{\alpha\beta}{\rm d}x^\alpha {\rm d}x^\beta\,;\\
R&=& \frac{1}{\sqrt{V_6M_{10}^6}}\,.\label{redmetric}
\eea
that guarantees that we are in the four dimensional Einstein frame given by ${\rm d} s_4^2$. Here $V_6$ is the volume of the compact extra dimensions, and 
$M_{10}$ is the 10-dimensional Planck mass. As explained in \cite{HNsoon}, if we use variables defined in terms of the 10-dimensional Planck mass as 
above (as opposed to 4 dimensional Planck units), experimental constraints force us to take only $n=2$ large extra dimensions, and the other four are 
fixed at the $M_{10}$ scale.

The KKLT-type model in \cite{Hinterbichler:2010wu} has a superpotential (KKLT has $e^{+i|a|\varrho}$) \footnote{As explained in \cite{Hinterbichler:2010wu},
it is not hard to obtain a term with 
the negative sign in the superpotential even in the KKLT context \cite{Abe:2005rx}, moreover with a highly suppressed prefactor A. 
There the superpotential with $A=Ce^{-m_9c<S>}$ is obtained by including gluino 
condensation on an extra D9-brane with magnetic flux in the KKLT scenario, where $2\pi S=e^{-\phi}-ic_0$ is the dilaton modulus, $c=8\pi^2/N_9$
and the coupling function on the D9-brane is $1/g_{D9}^2=|m_9 {\rm Re}S-w_9{\rm Re T}|$, with $T=-i\varrho$.
Moreover, rather generally, as explained in the review \cite{Quevedo:1996sv},
by imposing T-duality invariance (and modular invariance) on gaugino condensation superpotentials obtained for tori compactifications  
with dilaton modulus $S$ 
and volume modulus $T$, one obtains generically superpotentials of the form $W(S,T)\sim \eta(iT)^{-6}\exp (-3S/8\pi b)$, with $\eta(x)$ the 
Dedekind eta function. At large volume Re $T\rightarrow \infty$, as we need here, we obtain $W\propto e^{\pi T/2}$,
with a coefficient which is again exponentially small in the dilaton modulus, $\exp(-3S/8\pi b)$, with $b$ a renormalization group factor of order
1.}
\be
W(\varrho)=W_0+Ae^{-i|a|\varrho}
\ee 
written in terms of a complex scalar $\varrho$ whose imaginary part is related to volume $V_4$ of 4-cycles in the compact space that can be wrapped by 
Euclidean D3-branes,
\be
\sigma\equiv {\rm Im}\,\varrho=M_{10}^4V_4\sqrt{\pi}
\ee
The cycles that give the leading contribution are the largest ones, for which one finds
\be
\sigma= (M_{10})^2r^2\sqrt{\pi}=R^{-2}\sqrt{\pi}
\ee
and the factor of $\sqrt{\pi}$ can be absorbed in a trivial redefinition of parameters.

\begin{figure}
   \centering
   \includegraphics[width=4.0in]{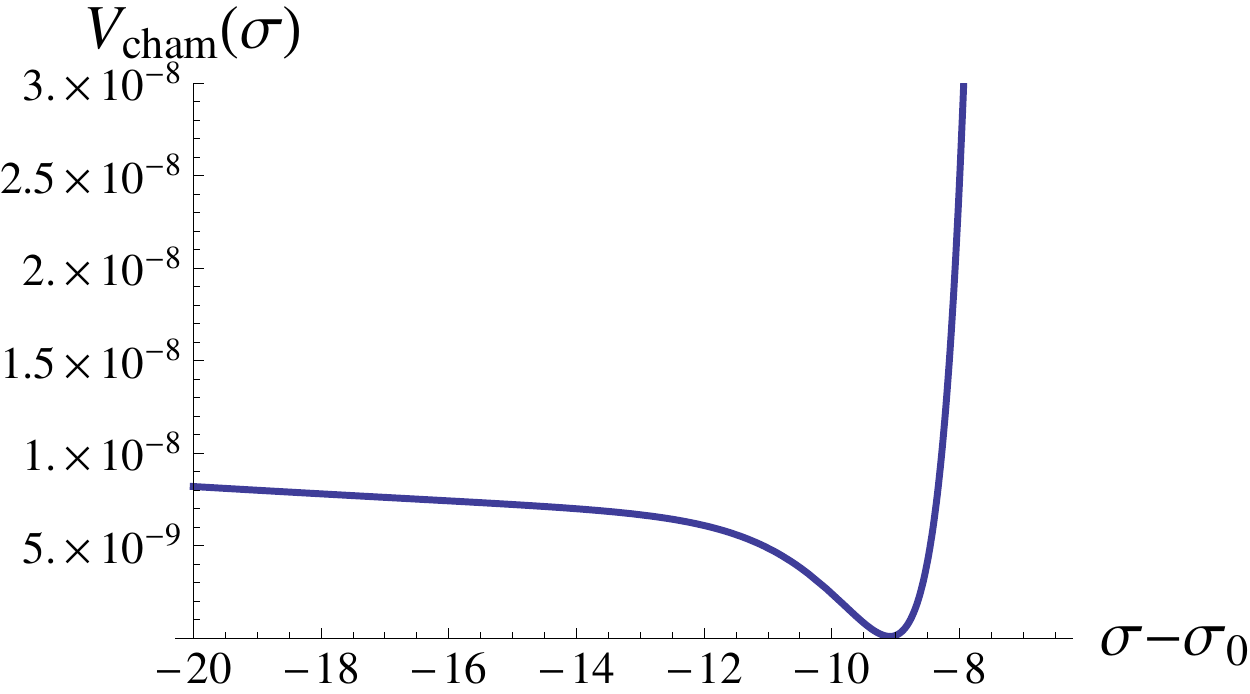}
   \caption{The chameleon potential as a function of $\sigma=R^{-2}$
   with $A\equiv e^{-|a|\sigma_0}$. For clarity of visualisation we plot numerically reasonable values of the parameters instead of realistic ones.}
      \label{champotential}
\end{figure}

The tree level K\"{a}hler potential for the case of $n=2$ is\footnote{For $n=6$ extra dimensions we have 
\be
K=-3M_{\rm Pl}^2\ln[-i(\rho-\bar\rho)]
\ee
but in general we write the reduction ansatz for the gravity action with an overall scale $\varrho$ for the extra dimensions, and find the K\"{a}hler
potential that gives the same scalar action. For $n=2$ we obtain the stated result.}
\be
K(\varrho,\bar\varrho)\simeq -2M_{\rm Pl}^2\ln[-i(\varrho-\bar\varrho)]\label{kahler}
\ee
The resulting supersymmetric potential is 
\be
V(R)=\frac{1}{2M_{\rm Pl}^2}\left[A^2a^2e^{2|a|R^{-2}}-2A|a|R^{2}e^{|a|R^{-2}}\Big(W_0+Ae^{|a|R^{-2}}\Big)-\frac{1}{2}R^{4}\Big(W_0+Ae^{|a|R^{-2}}\Big)^2\right]
\label{susypot}
\ee
and has a local AdS minimum at 
\be
\sigma_{\rm min} =R_{\rm min}^{-2} \approx \frac{1}{|a|} \ln\frac{W_0}{A}
\ee
As in KKLT, at the end we introduce a stack of antibranes, which break supersymmetry and adds a term $+D/\sigma^2$ to the potential.\footnote{Note that in KKLT one 
has a potential $D/\sigma^3$, which arises for $n=6$ large extra dimensions as follows \cite{Kachru:2002gs}. 
The volume modulus is written as $\varrho=ie^{4u}$, and then the 4d part of the metric is $g_{\mu\nu}^{(4)}=e^{-6u}\tilde g_{\mu\nu}^{(4)}$, 
with $\tilde g_{\mu\nu}^{(4)}$ the Einstein frame metric, leading to $\int d^4x\sqrt{-\det g_{\mu\nu}^{(4)}}=\int d^4x\sqrt{-\det \tilde g_{\mu\nu}^{(4)}}e^{-12u}
\propto 1/{\rm Im}\varrho^3=1/\sigma^3$. But for general $n$, $g_{\mu\nu}^{(4)}=e^{-nu}\tilde g_{\mu\nu}^{(4)}$, and for $n=2$ we have ${\rm Im}\varrho=ie^{2u}$,
and hence for $n=2$ we have $\int d^4x\sqrt{-\det g_{\mu\nu}^{(4)}}=\int d^4x\sqrt{-\det \tilde g_{\mu\nu}^{(4)}}e^{-4u}
\propto 1/{\rm Im}\varrho^2=1/\sigma^2$.} This term turns the local AdS minimum into a global dS minimum, as required by the observed cosmological constant.
The resulting potential is plotted in Fig.\ref{champotential} for values of the parameters which allow for visualization (instead of realistic ones).

The potential (\ref{susypot}) has a local minimum, around which we can make a quadratic approximation for $R>R_*$, and at some $R<R_*$ we can approximate it 
by the leading exponential, $\sim e^{2|a|R^{-2}}$. However, because the minimum of the potential itself is made up from the same exponentials, 
we cannot treat the region around the minimum as independent from the region $R<R_*$.

The constraint for laboratory experiments was found \cite{Hinterbichler:2010wu,HNsoon} to be expressed in terms of $R_*$ (here $R=R_*$ implies $\sigma=\sigma_*$), 
\be
\sigma_*=R_*^{-2}\gsim \frac{10^{30}}{|a|}\,,
\label{labconstr}
\ee
which, together with accelerator constraints on $M_{10}$ and gravity constraints on $r$ can only by satisfied by $n=2$ large extra dimensions
(in which case we have the $\sigma=R^{-2}$ as we have assumed until now). More precisely, we have now
\bea
M_{10}&\lsim & 2.5|a|^{1/2}~{\rm TeV}\,; \cr
r_* & \gsim & \frac{100}{|a|}~\mu{\rm m} \label{mrconstr}
\eea
We will consider in these constraints the natural value $|a|\sim 1$.
To satisfy collider constraints (for $n\neq 2$, from the equivalent of (\ref{labconstr}) we would find either a $M_{10}$ too low such that it would have been 
observed at current accelerators, or $r_*$ too high so that gravity experiments would have seen this effect, and for $n=2$ (\ref{mrconstr}) are only satisfied in a
braneworld scenario; see \cite{HNsoon} for more details), we then must have the Standard Model fields be confined to a brane, 
with the two large extra dimensions transverse to it, i.e. a $Dp$-brane with $p\leq 7$, situated at a fixed point in the extra dimensions. 

The K\"{a}hler potential (\ref{kahler}) implies that the canonical scalar field asssociated with $\varrho=i\sigma$ is 
\be
\phi=-M_{\rm Pl}\ln \frac{\sigma}{\sigma_*}\,,
\ee
where we have put $\phi=0$ at the present time, leading to a coupling function for the coupling to matter density (from (\ref{redmetric}))
\footnote{Note that the coupling here is fixed by the theory and is not a free parameter to be fixed by experiment as is usually assumed, see for example\cite{Brax:2010uq, Upadhye:2009iv, Steffen:2010ze}.}
\be
A(\phi) \equiv \frac{R}{R_*} = e^{\phi/2M_{\rm Pl}}\, .
\ee
That means that the effective potential, including the coupling to matter density $\rho$ is
\be
V_{\rm eff}(\phi)=V(\phi) + \rho \frac{R}{R_*} = V(\phi)+\rho e^{\phi/2M_{\rm Pl}}\,.\label{effpot}
\ee

While a potential which can be approximated by a quadratic around the minimum up to $R=R_*$ was found in \cite{Hinterbichler:2010wu} to 
constrain the mass of the chameleon on cosmological scales as $m_{\rm cosmo}\gsim 10^{3}H_0$, for the potential (\ref{susypot}) we have
$m_{\rm cosmo}\gsim 10^{15}H_0$. 

To avoid this stringent constraint, we now consider a racetrack type of superpotential \cite{Denef:2004dm}, with two exponentials in the superpotential instead of one, i.e.
\be
W(\varrho)=W_0+A_1e^{ia_1\varrho}+A_2e^{ia_2\varrho}\,,
\ee 
with $a_1,a_2<0$ and comparable, and $A_1,A_2>0$ and comparable as well, 
leading to a potential 
\bea
V(R)&=&\frac{1}{2M_{\rm Pl}^2}
\left[\left(A_1a_1e^{|a_1|R^{-2}}+A_2a_2e^{|a_2|R^{-2}}\right)^2-2R^2\left(A_1|a_1|e^{|a_1|R^{-2}}+A_2a_2e^{|a_2|R^{-2}}\right)
\times\right.\cr
&&\left.\times \Big(W_0+A_1e^{|a_1|R^{-2}}\Big)-\frac{1}{2}R^{4}\Big(W_0+A_1e^{|a_1|R^{-2}}+A_2e^{|a_2|R^{-2}}\Big)^2\right]\label{racepot}
\eea
We want the minimum and the leading exponential to be independent, so we need to choose the two exponentials to be very close, and $a_1$ to dominate 
the $\sigma>\sigma_*$ (or $R<R_*$) behaviour, while $a_2$ to dominate the minimum. In other words, we need $|a_1|>|a_2|$, yet very close in value, 
but we also need 
\be
A_2|a_2|e^{|a_2|\sigma_{\rm min}}>A_1|a_1|e^{|a_1|\sigma_{\rm min}}\label{expcond}
\ee
The condition for the minimum is $D_\varrho W=\d_\varrho W+(\d_\varrho K)W=0$, which gives 
\be
A_1|a_1|\sigma_{\rm min} e^{|a_1|\sigma}+A_2|a_2|\sigma_{\rm min} e^{|a_2|\sigma}=W_0+A_1e^{|a_1|\sigma}+A_2e^{|a_2|\sigma}
\ee
which can be solved (with the above assumptions, and considering that $|a_i|\sigma_{\rm min}\gg 1$) by
\be
\sigma_{\rm min}\simeq \frac{1}{|a_1|}\ln \frac{W_0}{|a_1|}\label{min}
\ee
and for which the minimum is 
\be
V_0\simeq -\frac{3W_0^2}{4\sigma_{\rm min}^2M_{\rm Pl}^2}\label{minvalue}
\ee

By simply allowing for this kind of superpotential we are allowed more freedom to fix our parameters and thus find a successful embedding. However, we should be cautious of such tuning as such as was pointed out in \cite{Greene:2005rn} such potentials are not necessarily stable to all corrections. Our goal here is simply to prove in principle that such an embedding can be done.

\section{Experimental constraints}

We already mentioned the constraint (\ref{labconstr}) obtained from laboratory experiments. A weaker constraint appears from the fact that the Milky Way 
galaxy must be screened, otherwise the field value
in the solar system, $\phi_{{\rm solar}\; {\rm system}}$, would not be fixed by the local density. That means that the galaxy needs to have a thin shell, i.e.
\be
\left(\frac{3\Delta {\cal R}}{{\cal R}}\right)_{\rm galaxy}  = \frac{\phi_{\rm cosmo} - \phi_{{\rm solar}\; {\rm system}}}{2gM_{\rm Pl}\Phi_{\rm G}} < 1\,,
\label{galaxyscreened}
\ee
where $\Phi_G\sim 10^{-6}$ is the Newtonian potential of the galaxy and $g=1/2$ is the chameleon coupling (see (\ref{effpot})). Since the field variations are 
small, we have $|\Delta \sigma/\sigma|\simeq 2\Delta R/R=\Delta \phi/M_{\rm Pl}$, meaning we obtain the constraint 
\be
\frac{\sigma_{\rm min} - \sigma_*}{\sigma_{\rm min}}\simeq
\frac{R_{\rm min} - R_{\rm solar\; system}}{2R_*} \;\lsim\; 10^{-6}\,,
\label{rconstra}
\ee
We will see later that this constraint is much weaker than (\ref{labconstr}), but constrains the same quantity. 

We want to derive a constraint on the mass of the chameleon field on cosmological scales, when the chameleon is close to the minimum. Therefore we want to find the constraint on 
\be
m^2_{\rm cosmo}\simeq \frac{g^2R_*^2}{M_{\rm Pl}^2} \left.\frac{{\rm d}^2V}{{\rm d}R^2}\right\vert_{R = R_{\rm min}}=
\frac{g^2R_*^2}{M_{\rm Pl}^2} \left[\frac{d\sigma}{dR}(R_{\rm min})\right]^2\frac{d^2V}{d\sigma^2}(\sigma_{\rm min})
\label{mmin}
\ee
Expanding the potential (\ref{racepot}) around the minimum (\ref{min}) with the value (\ref{minvalue}) for the potential, and 
assuming the condition (\ref{expcond}), we find that 
\be
\frac{d^2V}{d\sigma^2}(\sigma_{\rm min})\simeq \frac{4}{3}a_2^2|V_0|
\ee
Substituting back in (\ref{mmin}), and using $\sigma=R^{-2}$ and $g=1/2$, we get 
\be
m^2_{\rm cosmo}\simeq \frac{4}{3}a_2^2\sigma_{\rm min}^2\frac{|V_0|}{M_{\rm Pl}^2}\label{mcosmo}
\ee
so we see that a constraint on $m_{\rm cosmo}$ comes from a constraint on $|V_0|$. 

Such a constraint comes from imposing that the field value is not in the potential region of the leading exponential ($R<R_*$) in all the Universe, and that 
the low densities required to be on the quadratic region of the potential ($R>R_*$) are reached on some large scales, namely that the 
density $\rho_*$ corresponding to $R_*$ must be greater than the cosmic density, 
\be
\rho_*=R_*\left\vert\frac{{\rm d} V}{{\rm d}R}(R_*)\right\vert\geq H_0^2M_{\rm Pl}^2\,.
\label{rhoconstra}
\ee
But in the leading exponential region of the potential, 
\be
V(\sigma)-V_0\simeq \frac{A_1^2a_1^2}{2M_{\rm Pl}^2}e^{2|a_1|\sigma}
\ee
which gives
\be
\rho_*=R_*\left\vert\frac{d\sigma}{dR}(R_*)\frac{dV}{d\sigma}(\sigma_*)\right\vert \simeq \frac{2|a_1|\sigma_*A_1^2a_1^2}{M_{\rm Pl}^2}e^{2|a_1|\sigma_*}
\simeq \frac{8}{3}|a_1|\sigma_{\rm min}|V_0|e^{2|a_2|(\sigma_*-\sigma_{\rm min})}\geq H_0^2M_{\rm Pl}^2.\label{v0constr}
\ee
Since $a_1\simeq a_2$ and $\sigma_{\rm min}\simeq 10^{30}$, comparing (\ref{mcosmo}) with (\ref{v0constr}), we see that now we have an extra factor, 
$e^{|a_2|(\sigma_{\rm min}-\sigma_*)}$, to help. It was found in \cite{Hinterbichler:2010wu} that the constraint in the case of the phenomenological 
potential with an independent quadratic region near the minimum, and an unrelated leading exponential for $R<R_*$ was $m_{\rm cosmo}\geq 10^3H_0$. 
Since we could not do better, at most we can reach this constraint using the 
factor $e^{|a_2|(\sigma_{\rm min}-\sigma_*)}$. Direct comparison shows that a factor of $e^{|a_2|(\sigma_{\rm min}-\sigma_*)}=10^{12}$ would give us 
back the constraint $m_{\rm cosmo}\geq 10^3H_0$.

In the phenomenological potential, $\sigma_*$ (or $R_*$) is the separation point between the two (independent) parts of the potential, the quadratic and
leading exponential. In the case of our potential, we need a definition compatible with this phenomenological one.
Therefore we define $\sigma_*$ as the place where the derivative of the leading exponential equals the derivative of the other exponential, since this is 
indeeed the transition point from the leading exponential to the rest. We can check from (\ref{racepot}) that this gives approximately
\be
e^{(|a_1|-|a_2|)\sigma_*}=\frac{A_2a_2}{A_1a_1}
\ee
Calling by $K$ the ratio
\be
K=\frac{A_2|a_2|e^{|a_2|\sigma_{\rm min}}}{A_1|a_1|e^{|a_1|\sigma_{\rm min}}}
\ee
which had to be bigger than 1 according to (\ref{expcond}), the condition 
\be
e^{|a_2|(\sigma_{\rm min}-\sigma_*)}=10^{12}\label{ratio}
\ee
gives 
\be
\left(\frac{A_2|a_2|e^{|a_2|\sigma_{\rm min}}}{A_1|a_1|e^{|a_1|\sigma_{\rm min}}}=\right)
K=10^{12\frac{(|a_1|-|a_2|)}{|a_2|}}
\ee

We now also note that we need
\be
\sigma_*-\sigma_{\rm min}\simeq \frac{27}{|a_2|}
\ee
so going back to the galaxy screening constraint (\ref{rconstra}), it now translates into 
\be
|a_2|\sigma_*\gsim 3\times 10^7
\ee
which is much weaker than (\ref{labconstr}).

From (\ref{v0constr}), with (\ref{ratio}), we get
\be
|V_0| \;\gsim\; \frac{10^{-144}}{|a_1|\sigma_{\rm min}}M_{\rm Pl}^4
\ee
The constraint on $|a_1|\sigma_{\rm min}$ goes the opposite way, but assuming it is saturated at $10^{30}$, we get
\be
|V_0| \;\gsim\; 10^{-174}M_{\rm Pl}^4
\ee
and then also (from (\ref{minvalue}))
\be
W_0\;\gsim\; 10^{-57}M_{\rm Pl}^3
\ee
The coefficients $A$ are again extremely small, but  writing $A_ie^{|a_i|\sigma} = M_{\rm Pl}^3e^{|a_i|(\sigma-\sigma^i_0)}$, 
we have $\sigma^i_0\sim -\log A_i/M_{\rm Pl}^3\sim 10^{30}$.

Finally, when adding the supersymmetry-breaking antibrane term $+D/\sigma^2$  to the potential, as in KKLT, we can fix the value of $\sigma$ as follows. 
For the $A$'s and $W_0$'s we took, the value of the supersymmetric potential today is close to $V_{\rm today}\sim -10^{-174}M_{\rm Pl}^4$, negligible 
compared to the observed positive cosmological constant, hence we can assume that all the cosmological constant term comes from the supersymmetry-breaking
term, giving $D/\sigma_*^2\sim 10^{-122}M_{\rm Pl}^4$ or, since $\sigma_*\sim 10^{30}$, $D\sim 10^{-62}M_{\rm Pl}^4$. 
\Comment{ In KKLT D depends on the number of D branes and the warp factor in the throat where they sit. What is the physical picture here given the 
difference in dimensions? }

\section{Conclusions}

In this paper we have considered a ''racetrack" type superpotential $W=W_0+A_1a^{ia_1\varrho}+A_2e^{ia_2\varrho}$ instead of the single KKLT exponential, in 
order to obtain a chameleon scalar from a string theory context, generalizing the work in \cite{Hinterbichler:2010wu}. We have also used the more natural
large extra dimensional scenario from \cite{HNsoon}, in order to have $a_1,a_2$ closer to what can be obtained in KKLT. The simple modification of the 
"racetrack" allowed us to avoid having a too large chameleon mass on cosmological scales, and we found that we can have $m_{\rm cosmo}\geq 10^3H_0$, which can 
have implications interesting for cosmology.

{\bf Acknowledgements} We would like to thank Kurt Hinterbichler, Justin Khoury and Rogerio Rosenfeld for discussions. 
The work of HN is supported in part by CNPQ grant 301219/2010-9.
This material is based upon work supported financially by the National Research Foundation.
Any opinion, findings and conclusions or recommendations expressed in this
material are those of the authors and therefore the NRF does not accept any liability in regard thereto.

\bibliography{modchampaper}
\bibliographystyle{utphys}

\end{document}